\documentclass[preprint,prd,aps,showpacs,showkeys,nofootinbib]{revtex4}
\usepackage{graphicx}
\usepackage{color}
\usepackage{amssymb}
\begin{document}
\title{Search for charged lepton flavor violation of vector mesons in the $\mathrm{U}(1)_{X} \mathrm{SSM}$}
\author{Xing-Xing Dong$^{1,2,3}$\footnote{dongxx@hbu.edu.cn},
Shu-Min Zhao$^{1,2,3}$\footnote{zhaosm@hbu.edu.cn},
Jia-Peng Huo$^{1,2,3}$,
Tong-Tong Wang$^{1,2,3}$,
Yi-Tong Wang$^{1,2,3}$,
Tai-Fu Feng$^{1,2,3,4}$\footnote{fengtf@hbu.edu.cn}}

\affiliation{$^1$ College of Physics Science $\&$ Technology, Hebei University, Baoding, 071002, China\\
$^2$ Hebei Key Laboratory of High-precision Computation and Application of Quantum Field Theory, 071002, China\\
$^3$ Hebei Research Center of the Basic Discipline for Computational Physics, Baoding, 071002, China\\
$^4$ Department of Physics, Chongqing University, Chongqing, 401331, China}

\begin{abstract}
Charged lepton flavor violation (CLFV) represents a clear new physics (NP) signal beyond the standard model (SM). In this work, we investigate the CLFV decays of vector mesons $V\rightarrow l_i^{\pm}l_j^{\mp}$ with $V\in\{\phi, J/\Psi, \Upsilon(1S), \Upsilon(2S),\Upsilon(3S)\}$ in the $\mathrm{U}(1)_{X} \mathrm{SSM}$. Considering the SM-like Higgs boson mass within $3\sigma$ region, we discuss the corresponding constraints on the relevant parameter space of the model, which indicate this model can produce significant contributions to such CLFV decays. From the numerical analyses, the main sensitive parameters and CLFV sources originate from the non-diagonal elements correspond to the initial and final generations of the leptons. And the branching ratios to these CLFV processes can easily reach the present experimental upper bounds. Therefore, searching for CLFV processes of vector mesons may be an effective channel to study new physics.
\end{abstract}


\maketitle

\section{Introduction}
The standard model(SM) is considered as a much mature theory. However, the neutrino oscillation experiments have convinced that neutrinos possess tiny masses and mix with each other\cite{neutrino1,neutrino2,neutrino3,neutrino4,neutrino5}, which indicate that the charged lepton flavor violation (CLFV) is strongly suppressed in the SM\cite{SMLFV}. Therefore, some new physical (NP) models beyond the SM, which can easily generate CLFV precesses, have emerged. If the CLFV signals are observed in the future experiments, it is obvious evidence of the NP beyond the SM.

On the base of the minimal supersymmetric standard model (MSSM)\cite{MSSM1,MSSM2,MSSM3,MSSM4}, $\mathrm{U}(1)_{X} \mathrm{SSM}$ extends the gauge symmetry group to $SU(3)_C\otimes{SU(2)_L}\otimes{U(1)_Y}\otimes{U(1)_{X}}$\cite{SARAH1,U(1)X2 SARAH,SARAH2}. In this model, right-handed neutrinos and three Higgs singlets are added to MSSM. The right-handed neutrinos not only produce tiny masses to light neutrinos through seesaw mechanism, but also provide a new dark matter candidate. Besides, through the additional singlet Higgs states and right-handed (s)neutrinos, this model alleviates the hierarchy problem that appears in the MSSM. As the singlet Higgs superfield, $\hat{S}$ obtains a non-zero vacuum expectation value (VEV) $(v_{S} / \sqrt{2})$ after spontaneous breaking. Then the effective $\mu_{\mathrm{eff}}=\mu+\lambda_{H} v_{S} / \sqrt{2}$ term generated from $\mu \hat{H}_{u} \hat{H}_{d}$ and $\lambda_{H} \hat{S} \hat{H}_{u} \hat{H}_{d}$ can relieve the $\mu$ problem existing in the MSSM.

In this work, we investigate the CLFV processes of vector mesons $V\rightarrow l_i^{\pm}l_j^{\mp}$ with $V\in\{\phi, J/\Psi, \Upsilon(1S), \Upsilon(2S),\Upsilon(3S)\}$ in the $\mathrm{U}(1)_{X} \mathrm{SSM}$. In TABLE \ref{tab1}, we show the latest experimental data for these CLFV processes at $90\%$ confidence level (CL)\cite{Phiemuexp,Jpsaiemuexp,Jpsaietauexp,Jpsaimutauexp,R(1S)ljliexp,R(2S3S)etaumutauexp,R(3S)emuexp}. The SND detector at the VEPP-2M $e^+e^-$ collider reported the branching fraction of $\phi\rightarrow e\mu$ in 2010. BESII amd BESIII experiments gave the branching ratios of $J/\Psi\rightarrow \mu\tau$ and $J/\Psi\rightarrow e\mu (e\tau)$ respectively, especially the upper limit of $Br(J/\Psi\rightarrow e\tau)$ was renewed and reached to $7.5 \times 10^{-8}$ in 2022. The upper limits of CLFV decays $\Upsilon(1S)\rightarrow l_il_j$ were collected by the Belle detector at the KEKB collider in 2022. The BABAR detector discussed the experimental upper bounds of $Br(\Upsilon(2S)(\Upsilon(3S))\rightarrow e\tau(\mu\tau))$, and further researched the experiment limits of $Br(\Upsilon(3S)\rightarrow e\mu)$ in 2022 at the SLAC PEP-II $e^+e^-$ collider. These processes have been discussed in various theoretical frameworks, such as MRSSM\cite{MRSSM}, MSSM with type I seesaw mechanism\cite{MSSM type I1,MSSM type I2}, $Z'$ models\cite{Z' model}, leptoquark, SUSY, TC2 models\cite{TC2 model} and so on. In our previous work, we investigate the CLFV processes of vector mesons in the BLMSSM\cite{mesondecayBLMSSM}, and some of the theoretical evaluations fit better with the experimental upper bounds, which provide references and guidance for this work. Considering the SM-like Higgs boson mass within experimental $3\sigma$ region\cite{Higgsmassexp1,Higgsmassexp2,Higgsmassexp3} and the constraints from some CLFV process, such as $\mu\rightarrow e\gamma$\cite{mutoerexp} and $Z\rightarrow e\mu$\cite{Ztoemuexp}, we discuss the corresponding constraints on the relevant parameter space of the $\mathrm{U}(1)_{X} \mathrm{SSM}$. Through detailed analyses of these CLFV processes of vector mesons, we hope to reveal some properties of high energy physics.
\begin{table}[t]
\caption{ \label{tab1} The latest experiment limits for the CLFV decay ratios of vector mesons.}
\footnotesize
\begin{tabular*}{150mm}{@{\extracolsep{\fill}}cc|cc|cc}
\toprule  CLFV process&Present limit&CLFV process&Present limit&CLFV process&Present limit\\
\hline
$\phi\rightarrow e\mu$ & $2.0 \times 10^{-6}$\cite{Phiemuexp}&$J/\Psi\rightarrow e\tau$ & $7.5 \times 10^{-8}$\cite{Jpsaietauexp}&$J/\Psi\rightarrow \mu\tau$ & $2.0 \times 10^{-6}$\cite{Jpsaimutauexp}\\
$J/\Psi\rightarrow e\mu$ & $1.6 \times 10^{-7}$\cite{Jpsaiemuexp}&$\Upsilon(1S)\rightarrow e\tau$  &$2.7 \times 10^{-6}$\cite{R(1S)ljliexp}&$\Upsilon(1S)\rightarrow \mu\tau$  &$2.7 \times 10^{-6}$ \cite{R(1S)ljliexp}\\
$\Upsilon(1S)\rightarrow e\mu$  &$3.9 \times 10^{-7}$\cite{R(1S)ljliexp}&$\Upsilon(2S)\rightarrow e\tau$  &$3.2 \times 10^{-6}$\cite{R(2S3S)etaumutauexp}&$\Upsilon(2S)\rightarrow \mu\tau$  &$3.3 \times 10^{-6}$ \cite{R(2S3S)etaumutauexp}\\
$\Upsilon(3S)\rightarrow e\mu$  & $3.6 \times 10^{-7}$\cite{R(3S)emuexp}&$\Upsilon(3S)\rightarrow e\tau$  &$4.2 \times 10^{-6}$\cite{R(2S3S)etaumutauexp}&$\Upsilon(3S)\rightarrow \mu\tau$  &$3.1 \times 10^{-6}$\cite{R(2S3S)etaumutauexp} \\
\hline
\end{tabular*}%
\end{table}

This work is organized as follows. In Sec.II, we introduce the $\mathrm{U}(1)_{X} \mathrm{SSM}$ briefly. In Sec.III, we derive the analytic expressions for the CLFV ratios of vector mesons in the $\mathrm{U}(1)_{X} \mathrm{SSM}$. The numerical analyses are given out in Sec.IV, and the conclusion is discussed in Sec.V. The tedious couplings, hadron matrix elements and one-loop functions are collected in Appendix A, B and C, respectively.
\section{The $\mathrm{U}(1)_{X} \mathrm{SSM}$}
Adding the $\mathrm{U}(1)$ extension on the MSSM, the local gauge group of $\mathrm{U}(1)_{X} \mathrm{SSM}$ is $\mathrm{SU}(3)_{C} \otimes \mathrm{SU}(2)_{L} \otimes$ $\mathrm{U}(1)_{Y} \otimes \mathrm{U}(1)_{X}$. Comparing with MSSM, $\mathrm{U}(1)_{X} \mathrm{SSM}$ considers new superfields such as three Higgs singlets $\hat{\eta},\;\hat{\bar{\eta}},\;\hat{S}$ and right-handed neutrinos $\hat{\nu}_i$. The three light neutrinos can obtain tiny masses at tree level through the seesaw mechanism. The neutral CP-even Higgs $\phi_d^0,\;\phi_u^0,\;\phi_\eta^0,\;\phi_{\bar{\eta}}^0,\;\phi_S^0$ mix together, and form a $5 \times 5$ mass squared matrix. Besides, the lightest CP-even Higgs mass can be improved at tree level due to the introduction of $\eta,\;\bar{\eta},\;S$. The sneutrinos are departed into CP-even sneutrinos and CP-odd sneutrinos, whose mass squared matrices are both $6 \times6$. In TABLE \ref{tab2}, we show the chiral superfields and quantum numbers in the $\mathrm{U}(1)_{X} \mathrm{SSM}$.
Then the superpotential of $\mathrm{U}(1)_{X} \mathrm{SSM}$ is deduced as
\begin{eqnarray}
&&W= l_{W} \hat{S}+\mu \hat{H}_{u} \hat{H}_{d}+M_{S} \hat{S} \hat{S}-Y_{d} \hat{d} \hat{q} \hat{H}_{d}-Y_{e} \hat{e} \hat{l} \hat{H}_{d}+\lambda_{H} \hat{S} \hat{H}_{u} \hat{H}_{d} \nonumber \\
&&\hspace{0.6cm}+\lambda_{C} \hat{S} \hat{\eta} \hat{\bar{\eta}}+\frac{\kappa}{3} \hat{S} \hat{S} \hat{S}+Y_{u} \hat{u} \hat{q} \hat{H}_{u}+Y_{X} \hat{\nu} \hat{\bar{\eta}} \hat{\nu}+Y_{\nu} \hat{\nu} \hat{l} \hat{H}_{u}.
\end{eqnarray}
where $l_W$ is the parameter with mass squared dimension, $\mu$ and $M_S$ are both the parameters with mass dimension. $\mu$ indicates the supersymmetric mass between $SU(2)_{L}$ Higgs doublets $\hat{H}_d$ and $\hat{H}_u$, as well as $M_S$ represents the supersymmetric mass of $U(1)_{X}$ Higgs singlet $\hat{S}$.
\begin{table}[t]
\caption{ \label{tab2}  The chiral superfields and corresponding quantum numbers in the $\mathrm{U}(1)_{X} \mathrm{SSM}$.}
\begin{tabular*}{120mm}{@{\extracolsep{\fill}}cccccccccccc}
\toprule
Superfields & $\hat{q}_i$&$\hat{u}_i^c$ &$\hat{d}_i^c$ &$\hat{l}_i$ &$\hat{e}_i^c$ &$\hat{\nu}_i^c$ &$\hat{H}_u$ &$\hat{H}_d$ &$\hat{\eta}$&$\hat{\bar{\eta}}$&$\hat{S}$  \\
\hline
$SU(3)_C$&3 &$\bar{3}$ &$\bar{3} $&1 &1 &1 &1 &1 &1 &1 &1  \\

$SU(2)_L$&2 &1 &1 &2 &1 &1 &2 &2 &1 &1 &1  \\

$U(1)_Y$&1/6 &-2/3 &1/3 &-1/2 &1 &0 &1/2 &-1/2 &0 &0 &0  \\

$U(1)_X$ &0 &-1/2 &1/2 &0 &1/2 &-1/2 &1/2 &-1/2 &-1 &1 &0 \\
\hline
\end{tabular*}%
\end{table}

We show the concrete forms of the two Higgs doublets and three Higgs singlets
\begin{eqnarray}
&&H_{u}  =\left(\begin{array}{c}
H_{u}^{+} \\
\frac{1}{\sqrt{2}}\left(v_{u}+\phi_{u}^{0}+i P_{u}^{0}\right)
\end{array}\right),\hspace{0.5cm}  H_{d}=\left(\begin{array}{c}
\frac{1}{\sqrt{2}}\left(v_{d}+\phi_{d}^{0}+i P_{d}^{0}\right) \\
H_{d}^{-}
\end{array}\right),  \nonumber \\
&&\eta  =\frac{1}{\sqrt{2}}\left(v_{\eta}+\phi_{\eta}^{0}+i P_{\eta}^{0}\right),
\hspace{1.5cm}\bar{\eta}=\frac{1}{\sqrt{2}}\left(v_{\bar{\eta}}+\phi_{\bar{\eta}}^{0}+i P_{\bar{\eta}}^{0}\right), \nonumber \\
&&S  =\frac{1}{\sqrt{2}}\left(v_{S}+\phi_{S}^{0}+i P_{S}^{0}\right) .
\end{eqnarray}
Here, $v_{u}, v_{d}, v_{\eta}, v_{\bar{\eta}}$ and $v_{S}$ represent the VEVs of the Higgs superfields $H_{u}, H_{d}, \eta, \bar{\eta}$ and $S$, respectively. We define angle $\tan \beta_{\eta}=v_{\bar{\eta}} / v_{\eta}$ in analogy to the angle definition $\tan\beta=\frac{v_u}{v_d}$ in the MSSM.

The soft SUSY breaking terms in the $\mathrm{U}(1)_{X} \mathrm{SSM}$ are
\begin{eqnarray}
&&\mathcal{L}_{\mathrm{soft}}=\mathcal{L}_{\mathrm{soft}}^{\mathrm{MSSM}}-m_{\eta}^{2}|\eta|^{2}-m_{\bar{\eta}}^{2}|\bar{\eta}|^{2}
-m_{S}^{2} S^{2}-\left(m_{\tilde{\nu}_{R}}^{2}\right)^{I J} \tilde{\nu}_{R}^{I *} \tilde{\nu}_{R}^{J}\nonumber \\
&&\hspace{0.9cm}-B_{S} S^{2}-L_{S} S-\frac{T_{\kappa}}{3} S^{3}-T_{\lambda_{C}} S \eta \bar{\eta}+\epsilon_{i j} T_{\lambda_{H}} S H_{d}^{i} H_{u}^{j}-T_{X}^{I J} \bar{\eta} \tilde{\nu}_{R}^{* I} \tilde{\nu}_{R}^{* J}\nonumber \\
&&\hspace{0.9cm}-\epsilon_{i j} T_{\nu}^{I J} H_{u}^{i} \tilde{\nu}_{R}^{I *} \tilde{l}_{j}^{J}-\frac{1}{2}\left(M_{S} \lambda_{\tilde{X}}^{2}+2 M_{B B^{\prime}} \lambda_{\tilde{B}} \lambda_{\tilde{X}}\right)+\text { h.c. }
\end{eqnarray}
where $\lambda_{\tilde{B}}$ and $\lambda_{\tilde{X}}$ are the gauginos of $U(1)_Y$ and $U(1)_{X}$ respectively. Besides, we add the mass squared terms of sneutrinos and Higgs bosons, the trilinear scalar coupling terms and the Majorana mass terms to the soft breaking terms of the $\mathrm{U}(1)_{X} \mathrm{SSM}$.

Comparing with the MSSM, the two Abelian groups $\mathrm{U}(1)_{Y}$ and $\mathrm{U}(1)_{X}$ in the $\mathrm{U}(1)_{X} \mathrm{SSM}$ produce a new effect called as the gauge kinetic mixing, which can be induced through RGEs even with zero value at Grand Unified energy scale $M_{\mathrm{GUT}}$. In the general form, the covariant derivative of $\mathrm{U}(1)_{X} \mathrm{SSM}$ read as \cite{Dmu1,Dmu2,Dmu3}
\begin{eqnarray}
D_{\mu}=\partial_{\mu}-i\left(Y^{Y}, Y^{X}\right)\left(\begin{array}{cc}
g_{Y}, & g'_{Y X} \\
g'_{X Y}, & g'_{X}
\end{array}\right)\left(\begin{array}{c}
A_{\mu}^{'Y} \\
A_{\mu}^{'X}
\end{array}\right) ,
\end{eqnarray}
where $Y^{Y}$ denotes the $\mathrm{U}(1)_{Y}$ charge and $Y^{X}$ represents the $\mathrm{U}(1)_{X}$ charge, as well as $A_{\mu}^{'Y}$ and $A_{\mu}^{'X}$ are the gauge fields of $\mathrm{U}(1)_{Y}$ and $\mathrm{U}(1)_{X}$. We have proven that $\mathrm{U}(1)_{X} \mathrm{SSM}$ is anomaly free\cite{U(1)X anomaly free}. Then, we use the matrix $R$ to obtain coupling matrix with the two Abelian gauge groups unbroken condition
$\left(\begin{array}{cc}
g_{Y}, & g_{Y X}^{\prime} \\
g_{X Y}^{\prime}, & g_{X}^{\prime}
\end{array}\right) R^{T}=\left(\begin{array}{cc}
g_{1}, & g_{Y X} \\
0, & g_{X}
\end{array}\right) $.
Here, $g_1$ corresponds to the measured hypercharge coupling which is modified in $\mathrm{U}(1)_{X} \mathrm{SSM}$ as given along with $g_X$ and $g_{YX}$. Then, we can redefine the $U(1)$ gauge fields
$\left(\begin{array}{c}
A_{\mu}^{Y} \\
A_{\mu}^{X}
\end{array}\right)=R\left(\begin{array}{c}
A_{\mu}^{\prime Y} \\
A_{\mu}^{\prime X}
\end{array}\right)
$.

In this model, the gauge bosons $A_{\mu}^{X}, A_{\mu}^{Y}$ and $V_{\mu}^{3}$ mix together at the tree level, whose mass matrix is shown as follows
\begin{eqnarray}
\left(\begin{array}{ccc}
\frac{1}{8} g_{1}^{2} v^{2} & -\frac{1}{8} g_{1} g_{2} v^{2} & \frac{1}{8} g_{1}\left(g_{Y X}+g_{X}\right) v^{2} \\
-\frac{1}{8} g_{1} g_{2} v^{2} & \frac{1}{8} g_{2}^{2} v^{2} & -\frac{1}{8} g_{2}\left(g_{Y X}+g_{X}\right) v^{2} \\
\frac{1}{8} g_{1}\left(g_{Y X}+g_{X}\right) v^{2} & -\frac{1}{8} g_{2}\left(g_{Y X}+g_{X}\right) v^{2} & \frac{1}{8}\left(g_{Y X}+g_{X}\right)^{2} v^{2}+\frac{1}{8} g_{X}^{2} \xi^{2}
\end{array}\right),
\label{gaugemass}
\end{eqnarray}
with $v^{2}=v_{u}^{2}+v_{d}^{2}$ and $\xi^{2}=v_{\eta}^{2}+v_{\bar{\eta}}^{2}$. The mass matrix in Eq.(\ref{gaugemass}) can be diagonalized by Weinberg angle $\theta_{W}$ and the new mixing angle $\theta_{W}^{\prime}$. Here, $\sin ^{2} \theta_{W}^{\prime}$ is defined as
\begin{eqnarray}
\hspace{-0.5cm}&&\sin ^{2} \theta_{W}^{\prime}\nonumber \\
\hspace{-0.5cm}&&=\frac{1}{2}-\frac{\left(\left(g_{Y X}+g_{X}\right)^{2}-g_{1}^{2}-g_{2}^{2}\right) v^{2}+4 g_{X}^{2} \xi^{2}}{2 \sqrt{\left(\left(g_{Y X}+g_{X}\right)^{2}+g_{1}^{2}+g_{2}^{2}\right)^{2} v^{4}+8 g_{X}^{2}\left(\left(g_{Y X}+g_{X}\right)^{2}-g_{1}^{2}-g_{2}^{2}\right) v^{2} \xi^{2}+16 g_{X}^{4} \xi^{4}}} .
\end{eqnarray}
Then the eigenvalues of Eq.(\ref{gaugemass}) can be deduced as
\begin{eqnarray}
&&m_{\gamma}^{2}= 0, \nonumber \\
&&m_{Z, Z^{\prime}}^{2}= \frac{1}{8}\Big[\left(g_{1}^{2}+g_{2}^{2}+\left(g_{Y X}+g_{X}\right)^{2}\right) v^{2}+4 g_{X}^{2} \xi^{2}\nonumber \\
&&\mp \sqrt{\left(g_{1}^{2}+g_{2}^{2}+\left(g_{Y X}+g_{X}\right)^{2}\right)^{2} v^{4}+8\left(\left(g_{Y X}+g_{X}\right)^{2}-g_{1}^{2}-g_{2}^{2}\right) g_{X}^{2} v^{2} \xi^{2}+16 g_{X}^{4} \xi^{4}}\Big] .
\end{eqnarray}

We discuss the Higgs boson mass matrix because of the strict constraint from SM-like Higgs boson on the numerical results. The $\phi_d^0,\phi_u^0,\phi_\eta^0,\phi_{\bar{\eta}}^0,\phi_S^0$ mix together at the tree level and form the mass squared matrix for neutral CP-even Higgs boson $M_h^2$. The concrete form of the tree-level mass squared matrix can be generated by SARAH\cite{SARAH1,SARAH2}. $M_h^2$ is diagonalizd by $Z^H$, then we can derive the lightest tree-level Higgs boson mass $m_{h_1}^0$. Considering the leading-log radiative corrections from stop and top particles, the SM-like Higgs boson mass can be obtained \cite{leadinglog1,leadinglog2,leadinglog3}.
\begin{eqnarray}
m_h=\sqrt{(m_{h_1}^0)^2+\Delta m_h^2},
\end{eqnarray}
where the leading-log radiative corrections $\Delta m_h^2$ can be written as
\begin{eqnarray}
&&\Delta m_h^2=\frac{3m_t^4}{4 \pi ^2 v^2}\Big[\Big(\tilde{t}+\frac{1}{2}\tilde{X}_t\Big)+\frac{1}{16 \pi ^2}\Big(\frac{3m_t^2}{2v^2}-32\pi\alpha_3\Big)(\tilde{t}^2+\tilde{X}_t\tilde{t})\Big],\nonumber \\&&\tilde{t}=\log\frac{M_{SS}^2}{m_t^2},~~~~\tilde{X}_t=\frac{2\tilde{A}_t^2}{M_{SS}^2}(1-\frac{\tilde{A}_t^2}{12M_{SS}^2}).
\end{eqnarray}
Here, $\alpha_3$ is the strong coupling constant, $M_{SS}=\sqrt{m_{\tilde{t}_1}m_{\tilde{t}_2}}$ with stop masses $m_{\tilde{t}_{1,2}}$, trilinear Higgs-stops coupling $\tilde{A}_t=A_t-\mu  \cot \beta $ with $A_t=T_{u_{33}}$.

We also discuss some couplings used in this work. The corresponding contents can be given out in our Appendix A detailedly.
\section{The amplitudes for the CLFV decays of vector mesons}
In the $\mathrm{U}(1)_{X} \mathrm{SSM}$, we study the CLFV decays of vector mesons $V\rightarrow l_i^{\pm}l_j^{\mp}$ with $V\in\{\phi, J/\Psi, \Upsilon(1S), \Upsilon(2S),\Upsilon(3S)\}$. The relevant Feynman diagrams can be depicted in FIG.1, FIG.2 and FIG.3.
\begin{figure}[t]
\centering
\includegraphics[width=11cm]{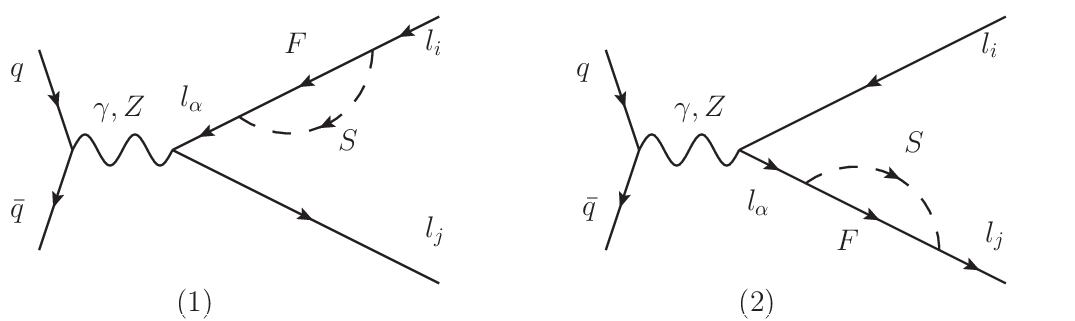}\\
\caption{The self-energy type diagrams for processes $V\rightarrow l_i^{\pm}l_j^{\mp}$, with $q$ representing $c$, $s$, $b$.} \label{fig3}
\end{figure}
\begin{figure}[t]
\centering
\includegraphics[width=11cm]{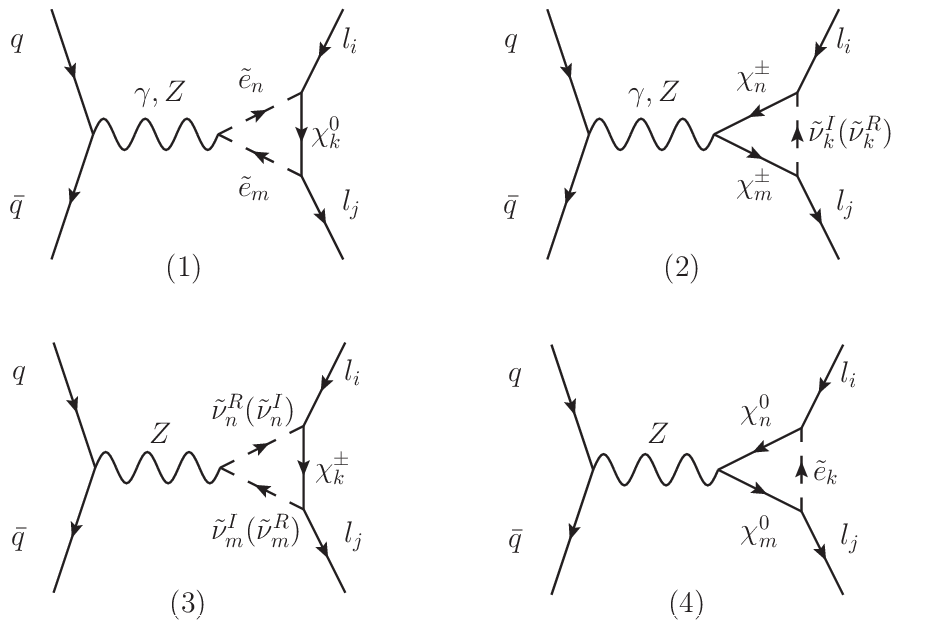}\\
\caption{The penguin type diagrams for processes $V\rightarrow l_i^{\pm}l_j^{\mp}$.} \label{fig1}
\end{figure}
\begin{figure}[t]
\centering
\includegraphics[width=11cm]{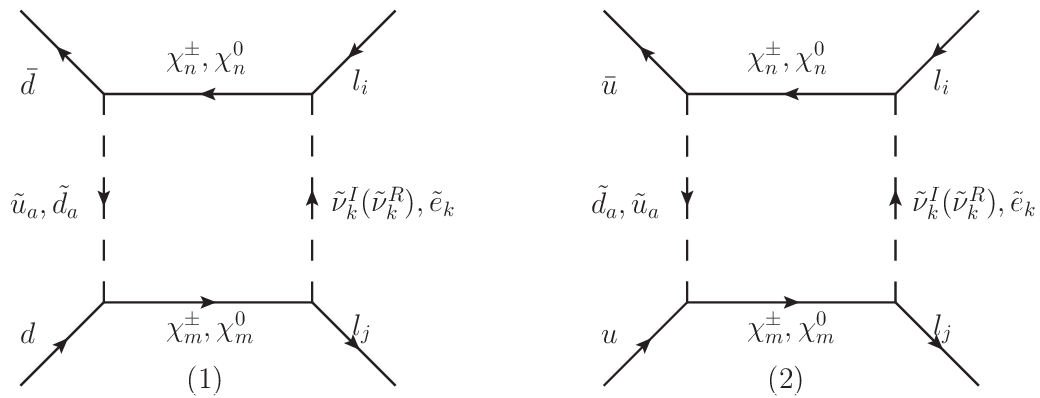}\\
\caption{The box type diagrams for processes $V\rightarrow l_i^{\pm}l_j^{\mp}$.} \label{fig2}
\end{figure}
In the quark picture, mesons are composed of a quark and an anti-quark, such as $\phi$ is made up of $s\bar{s}$, $J/\Psi$ is constituted of $c\bar{c}$ and $\Upsilon$ is composed of $b\bar{b}$. Here, a sum rule for light-cone wavefunction\cite{light-cone1,light-cone2,light-cone3,light-cone4,light-cone5,light-cone6,light-cone7} is adopted, which is widely used in the theoretical research of particle physics and nuclear physics.

We need to calculate the matrix elements of gauge invariant nonlocal operators $\langle0|\bar{q}(y)\Gamma[x,y]q(x)|V\rangle$ to obtain the decay amplitudes of processes involving the vector mesons\cite{light-cone4,light-cone5,light-cone6,light-cone7}. $\Gamma[x,y]$ is a generic Dirac matrix structure with $x$ and $y$ representing the coordinates of quark and anti-quark. Then we can deduce the leading-twist distribution amplitude of vector meson through the correlator\cite{light-cone4,light-cone5,light-cone6,light-cone7}:
\begin{eqnarray}
&&\langle0|\bar{q}_{\alpha}(y)q_{\beta}(x)|V(p)\rangle=\frac{\delta_{ij}}{4N_c}\int_0^1due^{-i(upy+\bar{u}px)}
[f_Vm_V{/\!\!\!\varepsilon}_VV_{||}(u)
\nonumber\\&&\hspace{4.2cm}+\frac{i}{2}\sigma^{\mu'\nu'}f_V^{\mathrm{\top}}(\varepsilon_{V\mu'}p_{\nu'}
-\varepsilon_{V\nu'}p_{\mu'})V_{\bot}(u)]_{\beta\alpha},
\end{eqnarray}
where the number of colors $N_c=3$. $m_V$ and $\varepsilon_V$ are respectively the mass and polarization vector of the vector meson. The vector meson momentum is on-shell, which indicates $p^2=m_V^2$. Here, the meson masses are adopted as $m_{\phi}=1.019$ GeV, $m_{J/\Psi}=3.097$ GeV, $m_{\Upsilon(1S)}=9.460$ GeV, $m_{\Upsilon(2S)}=10.023$ GeV and $m_{\Upsilon(3S)}=10.355$ GeV\cite{PDG2022,R(3S)mass}. $f_V$ and $f_V^{\top}$ are both the meson decay constants, for example, $f_{\phi}=0.24$ GeV, $f_{J/\Psi}=0.42$ GeV, $f_{\Upsilon(1S)}=0.65$ GeV, $f_{\Upsilon(2S)}=0.48$ GeV and $f_{\Upsilon(3S)}=0.54$ GeV, and we assume $f_V=f_V^{\top}$ in the follow calculation\cite{meson decay constant}. The integration variable $u$ and $\bar{u}\equiv1-u$ stand for the momentum fraction carried by the quark and anti-quark, respectively. $V_{||}(u)$($V_{\bot}(u)$) represents the leading-twist distribution function of longitudinal(transverse) polarized meson. The meson amplitudes are similar to their asymptotic forms\cite{asymptotic form}, so we make $V_{||}=V_{\bot}=V(u)=6u(1-u)$ in our calculation. Then the hadron matrix elements used in this work are discussed in Appendix B. In the frame of center of mass, the decay amplitudes of $V\rightarrow l_i^{\pm}l_j^{\mp}$ can be summarized at hadron level.
\subsection{The self-energy type diagrams}
We can generally write the effective amplitudes of $\gamma$ self-energy type diagrams corresponding to FIG.1(1) $\sim$ FIG.1(2):
\begin{eqnarray}
&&{\cal A}_{\gamma-s}={\bar u}_j(p_4)\gamma_{\mu}(A_LP_L+A_RP_R)v_i(p_3)\frac{e}{p^2}{\bar v}_V(p_2)H_V^{\gamma q\bar{q}}\gamma^{\mu}u_V(p_1),
\end{eqnarray}
where $p$ represents the meson momentum, $p_1$ and $p_2$($p_3$ and $p_4$) stand for the quark and anti-quark (lepton and anti-lepton) momentums respectively. ${\bar u}_j$ and $v_i$ are the wave functions of the $j$-th and $i$-th generation out-going leptons. As well as, ${\bar v}_V$ and $u_V$ are the wave functions of the incident quarks with $V\in\{\phi, J/\Psi, \Upsilon(1S), \Upsilon(2S), \Upsilon(3S)\}$. The coupling coefficients $H_V^{\gamma q\bar{q}}$ can be obtained from the vertexes given out in Appendix A, so as to the following $H_{L,R}^{\bar{l}_jFS}, H_{L,R}^{Zq\bar{q}}\cdot \cdot \cdot$. The Wilson coefficients $A_{L,R}$ from amplitudes of the FIG.1(1) and FIG.1(2) are shown as follows.
\begin{eqnarray}
&&A_L=\sum_{F=\chi^0,\chi^{\pm}}\sum_{S=\tilde{e},\tilde{\nu}^I(\tilde{\nu}^R)}\Big\{\frac{1}{x_{i}-x_{j}}
[x_i\sqrt{x_Fx_{j}}H_L^{\bar{l}_jFS}H_L^{l_i\bar{F}S^*}
\nonumber\\&&\hspace{1.2cm}-x_j\sqrt{x_Fx_{i}}H_R^{\bar{l}_jFS}H_R^{l_i\bar{F}S^*}][I_2(x_F,x_S)-I_3(x_F,x_S)]
-\frac{1}{2}H_R^{\bar{l}_jFS}H_L^{l_i\bar{F}S^*}I_5(x_F,x_S)\Big\},
\nonumber\\&&A_R=A_L|_{L{\leftrightarrow}R}.
\end{eqnarray}
Here, $x_{F,S}=m_{F,S}^2/\Lambda^2$, $m_{F,S}$ are the fermion and scalar boson mass, $\Lambda$ is the NP energy scale. The one-loop functions $I_i(x_1,x_2),i=1,2,\cdot\cdot\cdot,5$, as well as the following $G_i(x_1,x_2,x_3),\;i=1,2$ and $J_i(x_1,x_2,x_3,x_4),\;i=1,2$ are shown in Appendix C. Since the neutrino mass is very small, it makes little contribution to the problem we study, so the contributions to the self-energy type diagrams and penguin type diagrams from W boson-neutrino are ignore.

In the same way, the effective amplitudes of $Z$ self-energy type diagrams are deduced from FIG.1(1) and FIG.1(2).
\begin{eqnarray}
&&\hspace{-0.5cm}{\cal A}_{Z-s}={\bar u}_j(p_4)\gamma_{\mu}(B_LP_L+B_RP_R)v_i(p_3)\frac{1}{p^2-m_Z^2}{\bar v}_V(p_2)\gamma^{\mu}(H_L^{Zq\bar{q}}P_L+H_R^{Zq\bar{q}}P_R)u_V(p_1).
\end{eqnarray}
The concrete contributions from FIG. 1(1) and FIG. 1(2) are denoted by $B_{L,R}$.
\begin{eqnarray}
&&B_L=\hspace{-0.2cm}\sum_{F=\chi^0,\chi^{\pm}}\sum_{S=\tilde{e},\tilde{\nu}^I(\tilde{\nu}^R)}\hspace{-0.2cm}\Big\{\frac{1}{x_{i}-x_{j}}
[x_i\sqrt{x_Fx_{j}}H_L^{Z{l}_i\bar{l}_j}H_L^{\bar{l}_jFS}H_L^{l_i\bar{F}S^*}
\hspace{-0.2cm}-\hspace{-0.1cm}x_j\sqrt{x_Fx_{i}}H_L^{Z{l}_i\bar{l}_j}H_R^{\bar{l}_jFS}H_R^{l_i\bar{F}S^*}]
\nonumber\\&&\hspace{1.2cm}\times[I_2(x_F,x_S)-I_3(x_F,x_S)]
-\frac{1}{2}H_L^{Z{l}_i\bar{l}_j}H_R^{\bar{l}_jFS}H_L^{l_i\bar{F}S^*}I_5(x_F,x_S)\Big\},
\nonumber\\&&B_R=B_L|_{L{\leftrightarrow}R}.
\end{eqnarray}
\subsection{The penguin type diagrams}
We give out the penguin type diagrams for decays $V\rightarrow l_i^{\pm}l_j^{\mp}$ in FIG.2. The relevant effective amplitudes of $\gamma$ penguin type diagrams can be shown as
\begin{eqnarray}
&&{\cal A}_{\gamma-p}={\bar u}_j(p_4)[\gamma_{\mu}(C_1^LP_L+C_1^RP_R)+i\sigma_{\mu\nu}p^{\nu}(C_2^LP_L+C_2^RP_R)]v_i(p_3)
\nonumber\\&&\hspace{1.5cm}\times\frac{e}{p^2}{\bar v}_V(p_2)H_V^{\gamma q\bar{q}}\gamma^{\mu}u_V(p_1).
\end{eqnarray}
Firstly, we discuss the contributions $C_1^{L,R}(n),C_2^{L,R}(n)$ from the virtual neutral fermion diagram in the FIG.2(1).
\begin{eqnarray}
&&C_1^L(n)=\sum_{F=\chi^0}\sum_{S=\tilde{e}}
H_R^{SF\bar{l}_j}H_L^{S^*l_i\bar{F}}[\frac{1}{2}I_5(x_F,x_S)+\frac{p^2}{6\Lambda^2}I_4(x_F,x_S)],
\nonumber\\&&C_2^L(n)=\sum_{F=\chi^0}\sum_{S=\tilde{e}}\frac{m_F}{\Lambda^2}
H_L^{SF\bar{l}_j}H_L^{S^*l_i\bar{F}}[I_2(x_F,x_S)-I_3(x_F,x_S)],
\nonumber\\&&C_{\alpha}^R(n)=C_{\alpha}^L(n)|_{L{\leftrightarrow}R},\alpha=1,2.
\end{eqnarray}
Secondly, the couplings coefficients $C_1^{L,R}(c),C_2^{L,R}(c)$ correspond to the contributions from the virtual charged fermion diagram in the FIG.2(2), which are written as:
\begin{eqnarray}
&&C_1^L(c)=\sum_{F=\chi^{\pm}}\sum_{S=\tilde{\nu}^I(\tilde{\nu}^R)}
H_R^{SF\bar{l}_j}H_L^{S^*l_i\bar{F}}[\frac{1}{2}I_5(x_S,x_F)+x_FI_3(x_S,x_F)
\nonumber\\&&+\frac{p^2}{6\Lambda^2}(3I_2(x_S,x_F)-I_4(x_S,x_F))],
\nonumber\\&&C_2^L(c)=\sum_{F=\chi^{\pm}}\sum_{S=\tilde{\nu}^I(\tilde{\nu}^R)}\frac{m_F}{\Lambda^2}
H_L^{SF\bar{l}_j}H_L^{S^*l_i\bar{F}}I_2(x_S,x_F),
\nonumber\\&&C_{\alpha}^R(c)=C_{\alpha}^L(c)|_{L{\leftrightarrow}R},\alpha=1,2.
\end{eqnarray}

Then, the effective amplitudes of $Z$ penguin type diagrams are deduced in analogy to that of $\gamma$ penguin type diagrams. In addition, we just consider the dominate contributions of effective operators.
\begin{eqnarray}
\hspace{-0.4cm}&&{\cal A}_{Z-p}={\bar u}_j(p_4)\gamma_{\mu}(D_LP_L+D_RP_R)v_i(p_3)\frac{1}{p^2-m_Z^2}{\bar v}_V(p_2)\gamma^{\mu}(H_L^{Zq\bar{q}}P_L+H_R^{Zq\bar{q}}P_R)u_V(p_1).
\end{eqnarray}
The concrete contributions $D_{L,R}$ corresponding to FIG.2(1) $\sim$ FIG.2(4) are encoded as follows.
\begin{eqnarray}
&&D_L=-\frac{1}{2}\sum_{F=\chi^0,\chi^{\pm}}\sum_{S=\tilde{e},\tilde{\nu}^I(\tilde{\nu}^R)}
\big[-2\sqrt{x_{F_1}x_{F_2}}H_L^{SF_2\bar{l}_j}H_R^{ZF_1\bar{F}_2}H_L^{S^*l_i\bar{F}_1}G_1(x_S,x_{F_1},x_{F_2})
\nonumber\\&&+H_R^{SF_2\bar{l}_j}H_R^{ZF_1\bar{F}_2}H_L^{S^*l_i\bar{F}_1}G_2(x_S,x_{F_1},x_{F_2})
-H_R^{S_2F\bar{l}_j}H^{ZS_1S_2^*}H_L^{S_1^*l_i\bar{F}}G_2(x_F,x_{S_1},x_{S_2})\big],
\nonumber\\&&D_R=D_L|_{L{\leftrightarrow}R}.
\end{eqnarray}
\subsection{The box type diagrams}
In this subsection, we discuss the contributions of box type diagrams. In the calculation, we need to swap the position of the wave functions $u_V(p_1)$ and $v_i(p_3)$ to simplify the amplitudes. The method is named as Fierz Rearrangement, and the concrete transformation rules can be learnt from Refs.\cite{Fierz1,Fierz2,Fierz3}. Then the amplitudes can be deduced as:
\begin{eqnarray}
&&{\cal A}_{box}=\{N_1^L{\bar u}_j(p_4)\gamma_{\mu}P_Lv_i(p_3){\bar v}_V(p_2)\gamma^{\mu}P_Lu_V(p_1)+(L\leftrightarrow R)\}
\nonumber\\&&+\{N_2^L{\bar u}_j(p_4)\sigma_{\mu\nu}v_i(p_3){\bar v}_V(p_2)\sigma^{\mu\nu}P_Lu_V(p_1)+(L\leftrightarrow R)\}
\nonumber\\&&+\{N_3^L{\bar u}_j(p_4)\gamma_{\mu}P_Rv_i(p_3){\bar v}_V(p_2)\gamma^{\mu}P_Lu_V(p_1)+(L\leftrightarrow R)\},
\end{eqnarray}
where, the couplings coefficients $N_i^{L,R}$ correspond to the contributions from FIG.3.
\begin{eqnarray}
&&\hspace{-0.3cm}N_1^L=\frac{1}{4}\sum_{F_1,F_2=\chi^0,\chi^{\pm}}\sum_{S_1=\tilde{q}}\sum_{S_2=\tilde{e},\tilde{\nu}^I(\tilde{\nu}^R)}
H_L^{qS_1\bar{F}_2}H_R^{\bar{q}S_1^*F_1}H_L^{l_iS_2\bar{F}_1}H_R^{\bar{l}_jS_2^*F_2}J_1(x_{F_1},x_{F_2},x_{S_1},x_{S_2}),
\nonumber\\&&\hspace{-0.3cm}N_2^L=-\frac{\sqrt{x_{F_1}x_{F_2}}}{8}\sum_{F_1,F_2=\chi^0,\chi^{\pm}}\sum_{S_1=\tilde{q}}\sum_{S_2=\tilde{e},\tilde{\nu}^I(\tilde{\nu}^R)}
\hspace{-0.3cm}H_L^{qS_1\bar{F}_2}H_L^{\bar{q}S_1^*F_1}H_L^{l_iS_2\bar{F}_1}H_L^{\bar{l}_jS_2^*F_2}\hspace{-0.1cm}J_2(x_{F_1},x_{F_2},x_{S_1},x_{S_2}),
\nonumber\\&&\hspace{-0.3cm}N_3^L=-\frac{\sqrt{x_{F_1}x_{F_2}}}{2}\hspace{-0.3cm}\sum_{F_1,F_2=\chi^0,\chi^{\pm}}\hspace{-0.1cm}\sum_{S_1=\tilde{q}}\sum_{S_2=\tilde{e},\tilde{\nu}^I(\tilde{\nu}^R)}
\hspace{-0.4cm}H_L^{qS_1\bar{F}_2}H_R^{\bar{q}S_1^*F_1}H_R^{l_iS_2\bar{F}_1}H_L^{\bar{l}_jS_2^*F_2}\hspace{-0.1cm}J_2(x_{F_1},x_{F_2},x_{S_1},x_{S_2}).
\end{eqnarray}

The branching ratios for processes $V\rightarrow l_i^{\pm}l_j^{\mp}$ can be deduced as:
\begin{eqnarray}
&&Br\left(V\rightarrow l_i^{\pm}l_j^{\mp}\right)=\frac{\sqrt{[m_V^2-(m_{l_i}+m_{l_j})^2][m_V^2-(m_{l_i}-m_{l_j})^2]}}{16\pi m_V^3\Gamma_V}\times\sum_{\xi}{\cal A}_{V_{\xi}}{\cal A}_{V_{\xi}}^*,
\end{eqnarray}
where $\Gamma_V$ represents the total decay width of meson, for example $\Gamma_{\phi}\simeq4.2\times10^{-3}$ GeV, $\Gamma_{J/\Psi}\simeq0.0926\times10^{-3}$ GeV, $\Gamma_{\Upsilon(1S)}\simeq0.054\times10^{-3}$ GeV, $\Gamma_{\Upsilon(2S)}\simeq0.032\times10^{-3}$ GeV and $\Gamma_{\Upsilon(3S)}\simeq0.020\times10^{-3}$ GeV\cite{PDG2022}. $A_{V_{\xi}}$ are the amplitudes corresponding to FIG.1, FIG.2 and FIG.3.
\section{The numerical results}
In this section, we research the numerical results of the CLFV processes $V\rightarrow l_i^{\pm}l_j^{\mp}$ with $V\in\{\phi, J/\Psi, \Upsilon(1S), \Upsilon(2S), \Upsilon(3S)\}$. We consider some experimental constraints: 1. The updated experimental data indicates that the $Z'$ boson mass satisfies $M_{Z'}\geq 5.15~ {\rm TeV}$ with $95\%$ CL\cite{Zpupper}. The upper bound on the ratio between the $Z'$ boson mass and gauge coupling $g_X$ is approximatively smaller than $6~{\rm TeV}$ at $99\%$ CL\cite{Zpupper1,Zpupper2}, so $g_X$ is restricted in the interval of $0\sim0.85$. 2. The $\bar{B}\rightarrow X_s\gamma$ experiment has excluded the large $\tan\beta$\cite{BSgamma1,BSgamma2}. 3. $\tan\beta_\eta$ is smaller than 1.5 after taking into account the constraint from LHC data\cite{tbyita}. 4. According to the latest LHC data, the slepton mass is greater than 700 GeV, the chargino mass is greater than 1100 GeV and the squark mass is greater than 1500 GeV in our numerical discussions \cite{scalar mass exp1,scalar mass exp2,scalar mass exp3,scalar mass exp4,scalar mass exp5,scalar mass exp6}. 5. The SM-like Higgs boson mass, $m_h=125.25\pm0.17$ GeV, constrains the parameter space strictly\cite{PDG2022,Higgsmassexp1,Higgsmassexp2,Higgsmassexp3}. Therefore, we limit the SM-like Higgs boson mass of the $\mathrm{U}(1)_{X} \mathrm{SSM}$ within experimental $3\sigma$ region. 6. We consider the constraints from some CLFV processes, such as $Br(\mu\rightarrow e\gamma)<5.7\times10^{-13}$ and $Br(Z\rightarrow e\mu)<7.5\times10^{-7}$\cite{mutoerexp,Ztoemuexp}. Then the numerical results of the CLFV processes $V\rightarrow l_i^{\pm}l_j^{\mp}$ in the $\mathrm{U}(1)_{X} \mathrm{SSM}$ are discussed detailedly.

Considering the above experimental constraints, some parameters we used are shown as
\begin{eqnarray}
&& \lambda_{H}=\lambda_{C}=\kappa=0.1,\;\tan \beta_{\eta}=0.8,\;Y_{X_{ii}}=0.5,\;g_X=0.3,\;M_2=1.2\;\mathrm{TeV},
\nonumber\\&&M_{BB'}=0.4\;\mathrm{TeV},\;\mu=M_1=M_{BL}=T_{\lambda_{H}}=T_{\lambda_{C}}=T_{\kappa}=T_{u_{ii}}=T_{d_{ii}}=1 \;\mathrm{TeV},
\nonumber\\&&T_{X_{ii}}=-1\;\mathrm{TeV},\; T_{E_{ii}}=-3\;\mathrm{TeV},\; T_{\nu_{ii}}=0.5\;\mathrm{TeV},\;v_{S}=3 \;\mathrm{TeV},\;m_{{\nu}_{R_{ii}}}^{2}=0.5\;\mathrm{TeV}^{2},
\nonumber\\&& m_{{Q}_{ii}}^{2}=m_{{U}_{ii}}^{2}=m_{{D}_{ii}}^{2}=10\;\mathrm{TeV}^{2}, m_{S}^{2}=m_{{E}_{ii}}^{2}=1\;\mathrm{TeV}^{2},\;l_{W}=B_{\mu}=B_{S}=0.1\;\mathrm{TeV}^{2}.
\end{eqnarray}
\subsection{$V\rightarrow e^+\mu^-$}
\begin{figure}[t]
\centering
\includegraphics[width=7.5cm]{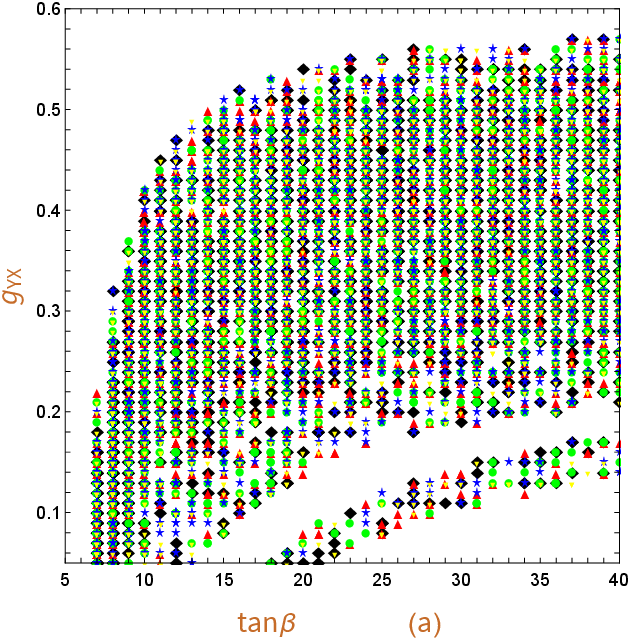}\hspace{0.5cm}
\includegraphics[width=7.6cm]{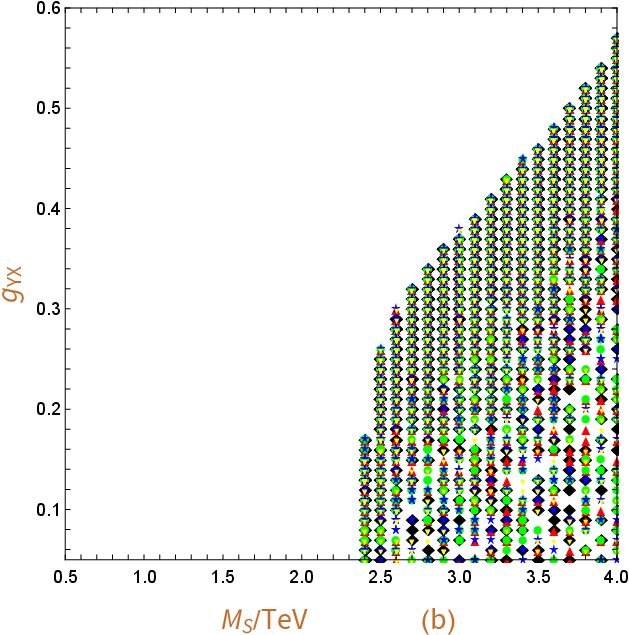}\\
\caption{$\tan\beta$ ($M_S$) varying with $g_{YX}$ are plotted, where the vector mesons of $\phi$, $J/\Psi$, $\Upsilon(1S)$, $\Upsilon(2S)$, $\Upsilon(3S)$ are represented by $\blacklozenge$, {\color{red} {$\blacktriangle$}}, {\color{green} {$\bullet$}}, {\color{blue} {$\bigstar$}}, {\color{yellow} {$\blacktriangledown$}} respectively.}
\label{fig4}
\end{figure}

Parameters $\tan\beta$, $g_{YX}$, $M_1$, $M_2$, $M_S$, $M_{BB'}$, $M_{BL}$, $m_{{L}_{ii}}^{2}$, $m_{{E}_{ii}}^{2}$, $m_{{\nu}_{R_{ii}}}^{2}$, $T_{X_{ii}}$, $T_{E_{ii}}$, $T_{\nu_{ii}}$ and $m_{{L}_{ij}}^{2}$, $m_{{E}_{ij}}^{2}$, $m_{{\nu}_{R_{ij}}}^{2}(i\neq j)$ are assumed as random variables in the suitable regions. As the SM-like Higgs boson mass is in $3\sigma$ region, and the branching ratios of $V\rightarrow e^+\mu^-$ satisfy the current experiment constraints, the reasonable parameter space is selected to scatter points, which are shown in FIG.\ref{fig4}. Firstly, we discuss the distribution of $\tan\beta$ versus $g_{YX}$ in FIG.\ref{fig4}(a). On the whole, the upper and lower limits of $g_{YX}$ almost increase with the enlarged $\tan\beta$. It is worth noting that when $\tan\beta$ changes from 7 to 10, the upper limit of $g_{YX}$ increases significantly, and then its upper limit increases slowly as $\tan\beta$ increases. And when $\tan\beta$ is between 14 and 40, $g_{YX}$ can fetch almost any value in the range of 0.2 to 0.5. FIG.\ref{fig1}(b) indicates that $M_S$ is larger than 2.4 TeV, as well as, the upper limit and the suitable parameter space of $g_{YX}$ both increase obviously with the increase of $M_S$. Therefore, in the numerical analyses below, we take $g_{YX}=0.3$, $\tan\beta=15$ and $M_S=2.7$ TeV.

\begin{figure}[t]
\centering
\includegraphics[width=12cm]{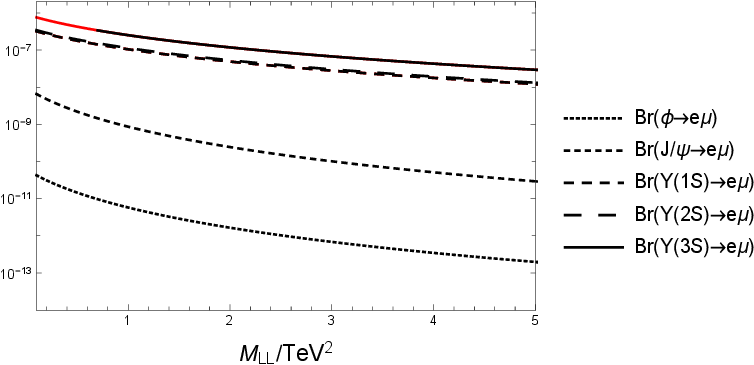}\\
\caption{The contributions to $V\rightarrow e^+\mu^-( V\in\{\phi, J/\Psi, \Upsilon(1S), \Upsilon(2S), \Upsilon(3S)\})$ varying with $M_{LL}$ are respectively plotted.} \label{fig5}
\end{figure}

\begin{figure}[t]
\centering
\includegraphics[width=12cm]{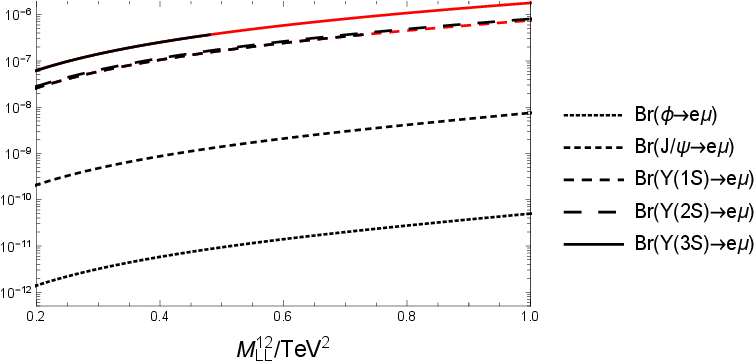}\\
\caption{The contributions to $V\rightarrow e^+\mu^-( V\in\{\phi, J/\Psi, \Upsilon(1S), \Upsilon(2S), \Upsilon(3S) \})$ varying with $M_{LL}^{12}$ are respectively plotted.} \label{fig6}
\end{figure}

The CLFV process $V\rightarrow e^+\mu^-$ is flavor dependent, which can be influenced by the parameters $M_{LL}=m_{{L}_{ii}}^{2}, M_{LL}^{ij}=m_{{L}_{ij}}^{2}, M_{EE}=m_{{E}_{ii}}^{2}, M_{EE}^{ij}=m_{{E}_{ij}}^{2},M_{NU}^{ij}=m_{{\nu}_{R_{ij}}}^{2}\cdot\cdot\cdot$. When $M_{LL}^{12}=0.4\;\mathrm{TeV}^2$ and $M_{EE}^{12}=M_{NU}^{12}=10^{-2}\;\mathrm{TeV}^2$, $Br(V\rightarrow e^+\mu^-)$ versus the parameter $M_{LL}$ are plotted in FIG.\ref{fig5}, as well as $Br(V\rightarrow e^+\mu^-)$ versus the parameter $M_{LL}^{12}$ are figured in FIG.\ref{fig6} as $M_{LL}=1\;\mathrm{TeV}^2$ and $M_{EE}^{12}=M_{NU}^{12}=10^{-2}\;\mathrm{TeV}^2$. The red line has been excluded due to exceed experimental constraints of SM-like Higgs mass $3\sigma$ region or $Br(V\rightarrow e^+\mu^-)$. It is obvious that the CLFV rates of $V\rightarrow e^+\mu^-$ all decrease with the enlarged $M_{LL}$, and increase with the enlarged $M_{LL}^{12}$. Moreover, the present  experimental upper bound of $Br(\Upsilon(3S)\rightarrow e^+\mu^-)$ possesses the most obvious constraints on parameter $M_{LL}$ and $M_{LL}^{12}$, which are $M_{LL}\geq0.7\;\mathrm{TeV}^2$ and $M_{LL}^{12}\leq0.45\;\mathrm{TeV}^2$. In the follow discussion, we take $M_{LL}=1\;\mathrm{TeV}^2$.
\subsection{$V\rightarrow e^+\tau^-$ and $V\rightarrow \mu^+\tau^-$}
\begin{figure}[t]
\centering
\includegraphics[width=12cm]{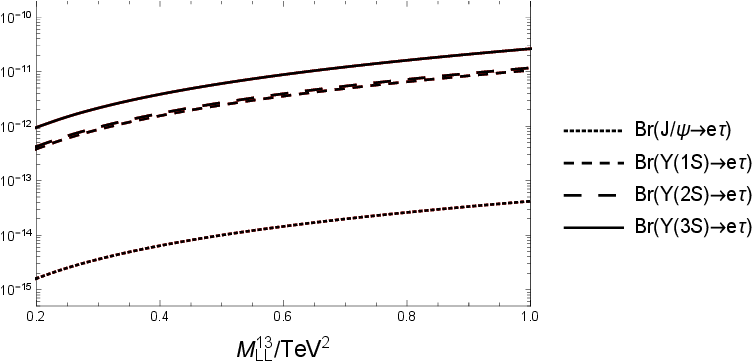}\\
\caption{The contributions to $V\rightarrow e^+\tau^-( V\in\{J/\Psi, \Upsilon(1S), \Upsilon(2S), \Upsilon(3S), \})$ varying with $M_{LL}^{13}$ are respectively plotted.} \label{fig7}
\end{figure}

\begin{figure}[t]
\centering
\includegraphics[width=9cm]{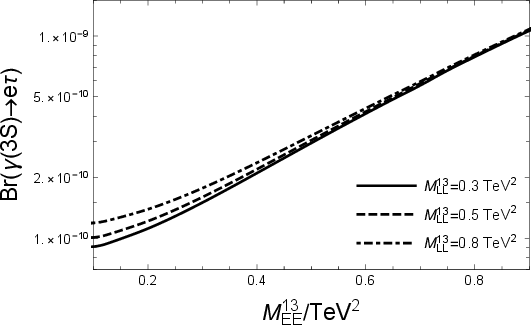}\\
\caption{The contributions to $\Upsilon(3S)\rightarrow e^+\tau^-$ varying with $M_{EE}^{13}$ are plotted, where the solid, dashed and dot-dashed lines correspond to $M_{LL}^{13}=0.3, 0.5,0.8\;\mathrm{TeV}^2$, respectively.} \label{fig8}
\end{figure}

In this subsection, we first study the CLFV rates for $V\rightarrow e^+\tau^-$ versus $M_{LL}^{13}$ and $M_{EE}^{13}$ respectively in FIG.\ref{fig7} and FIG.\ref{fig8}. In general, the numerical results of $Br(V\rightarrow e^+\tau^-)$ both enlarge with the increase of $M_{LL}^{13}$ and $M_{EE}^{13}$. Obviously, FIG.\ref{fig7} indicates that when $M_{EE}^{13}=M_{NU}^{13}=0$, the CLFV rates of the processes $V\rightarrow e^+\tau^-( V\in\{\Upsilon(1S), \Upsilon(2S), \Upsilon(3S)\})$ changing with $M_{LL}^{13}$ are very close, and nearly three orders of magnitude larger than the $Br(J/\Psi\rightarrow e^+\tau^-)$. Then, we take the process $ \Upsilon(3S)\rightarrow e^+\tau^-$ as an example, and discuss the CLFV ratio of this process with the parameter $M_{EE}^{13}$ in FIG.\ref{fig8}  as $M_{NU}^{13}=5\times10^{-2}\;\mathrm{TeV}^2$. We find that when the parameter $M_{LL}^{13}$ enlarges from $0.3\;\mathrm{TeV}^2$ to $0.8\;\mathrm{TeV}^2$, the numerical result increases slowly. However, the numerical results can be improved about an order of magnitude with the increase of $M_{EE}^{13}$, which indicates that the effect of parameter $M_{LL}^{13}$ on the numerical results is slightly smaller than that of parameter $M_{EE}^{13}$.

\begin{figure}[t]
\centering
\includegraphics[width=12cm]{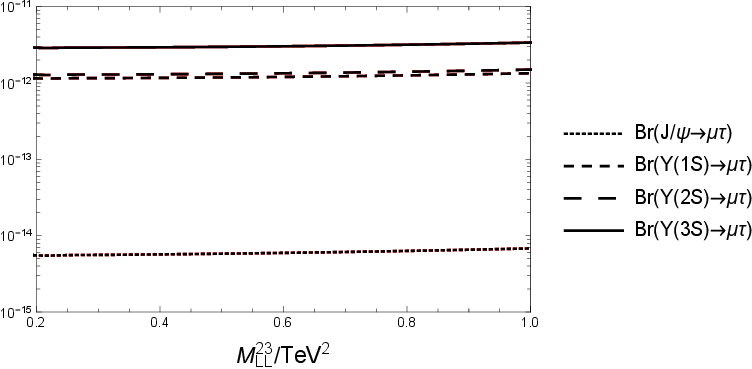}\\
\caption{The contributions to $V\rightarrow \mu^+\tau^-( V\in\{J/\Psi, \Upsilon(1S), \Upsilon(2S), \Upsilon(3S), \})$ varying with $M_{LL}^{23}$ are respectively plotted.} \label{fig9}
\end{figure}

\begin{figure}[t]
\centering
\includegraphics[width=9cm]{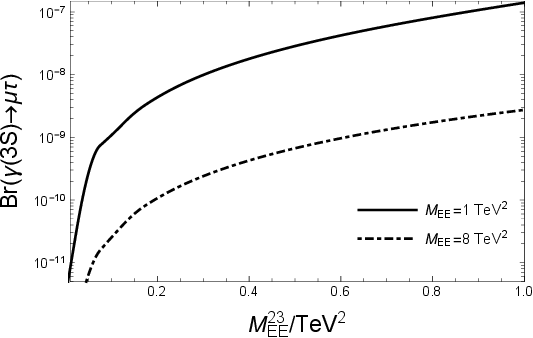}\\
\caption{The contributions to $\Upsilon(3S)\rightarrow \mu^+\tau^-$ varying with $M_{EE}^{23}$ are respectively plotted, where the solid and dot-dashed lines correspond to $M_{EE}=1,8\;\mathrm{TeV}^2$, respectively.} \label{fig10}
\end{figure}

When $M_{EE}^{23}=M_{NU}^{23}=0$, we analyze the transformations of the CLFV ratios of processes $V\rightarrow \mu^+\tau^-$ with the parameter $M_{LL}^{23}$, as shown in FIG.\ref{fig9}. By observing this figure, it is easy to find that although parameter $M_{LL}^{23}$ has the lifting effect on the numerical results, the effect is not obvious. Therefore, the non-diagonal element $M_{LL}^{23}$, from slepton matrix $m_L^2$, possesses weak influence on the CLFV decays. Taking the process $\Upsilon(3S)\rightarrow \mu^+\tau^-$ as an example, we discuss how the CLFV ratio of this process varies with the diagonal element $M_{EE}$ and the non-diagonal element $M_{EE}^{23}$ of the slepton matrix $m_{E}^2$ in FIG.\ref{fig10}, where $M_{LL}^{23}=0.5\;\mathrm{TeV}^2$ and $M_{NU}^{23}=0.05\;\mathrm{TeV}^2$. We find that an increase in diagonal element $M_{EE}$ significantly depresses the numerical result, while an increase in non-diagonal element $M_{EE}^{23}$ significantly increases the numerical result. Therefore, $M_{EE}$ and $M_{EE}^{23}$ are both the sensitive parameters.
\section{discussion and conclusion}
In this work, we focus on the CLFV decays of vector mesons $V\rightarrow l_i^{\pm}l_j^{\mp}$ with $V\in\{\phi, J/\Psi, \Upsilon(1S), \Upsilon(2S),\Upsilon(3S)\}$ in the $\mathrm{U}(1)_{X} \mathrm{SSM}$. We find that the $\mathrm{U}(1)_{X} \mathrm{SSM}$ contributions for CLFV processes beyond MSSM are considerable. In the numerical discussion, we constrain the SM-like Higgs mass within $3\sigma$ region, which possesses strict limit on the CLFV decays. Firstly, the distribution of $\tan\beta$ versus $g_{YX}$ indicates $g_{YX}$ can fetch almost any value in the range of 0.2 to 0.5 as $\tan\beta$ is between 14 and 40. Besides, $M_S$ is larger than 2.4 TeV, as well as, the upper limit and the suitable parameter space of $g_{YX}$ both increase obviously with the increase of $M_S$. Not only that, the branching ratios of $V\rightarrow l_i^{\pm}l_j^{\mp}$ depend on the slepton flavor mixing parameters $M_{LL}^{ij}$ and $M_{EE}^{ij}$ obviously, and increase with the enlarged $M_{LL}^{ij}$ and $M_{EE}^{ij}$. Considering the latest experimental limits of $Br(\Upsilon(3S)\rightarrow e^+\mu^-)$, the slepton flavor mixing parameter $M_{LL}^{12}$ is restricted as $M_{LL}^{12}\leq0.45 \;\mathrm{TeV}^2$. In addition, $M_{LL}$, as the diagonal element of the slepton and sneutrino, is limited as $M_{LL}\geq0.7\;\mathrm{TeV}^2$ by the latest experimental limits of $Br(\Upsilon(3S)\rightarrow e^+\mu^-)$. And the increase of $M_{LL}$ has the significant inhibitory effect on the numerical results.

After considering the SM-like Higgs mass within $3\sigma$ region, the experimental constraints on $Br(\mu\to e+\gamma)$, $Br(Z\to e\mu)$ and $Br(V\to e+\mu^-)$, we find that $Br(\phi\rightarrow e\mu)\sim10^{-11},\; Br(J/\Psi\rightarrow e\mu)\sim10^{-9}$ and $Br(\Upsilon(1S)(\Upsilon(2S),\Upsilon(3S))\rightarrow e\mu)\sim10^{-7}$. The decays $\Upsilon(1S)(\Upsilon(2S),\Upsilon(3S))\rightarrow e\mu$ are much easier than $\phi(J/\Psi)\rightarrow e\mu$ to reach the experimental upper bounds. Similarly, $Br(\Upsilon(3S)\rightarrow e\tau)$ are at the order of $10^{-9}$ and $Br(\Upsilon(3S)\rightarrow\mu\tau)$ can reach $10^{-7}$. Above results indicate that processes $\Upsilon\rightarrow e\mu$ and $\Upsilon(3S)\rightarrow \mu\tau$ are very promising to be observed in the near future experiments. A summary table of the theoretical predictions of $Br(J/\Psi(\Upsilon)\rightarrow e\mu)$ and $Br(\Upsilon(3S)\rightarrow \mu\tau)$ in literatures are presented in TABLE \ref{tab3}, which indicates that the corresponding theoretical predictions in our model may be realized much easilier than that in the SUSY, 331 and 211 models.
\begin{table}[t]
\caption{ \label{tab3}  The theoretical predictions of $Br(J/\Psi(\Upsilon)\rightarrow e\mu)$ and $Br(\Upsilon(3S)\rightarrow \mu\tau)$ in literatures.}
\begin{tabular*}{140mm}{@{\extracolsep{\fill}}ccccc}
\toprule
$Br(V \to l_il_j)$& $\mathrm{U}(1)_{X} \mathrm{SSM}$ &SUSY madel\cite{TC2 model} &331 model\cite{Z' model} & 211 model\cite{Z' model}  \\
\hline
$Br(J/\Psi\rightarrow e\mu)$&$\sim10^{-9}$ &$\sim10^{-16}$  &$\sim10^{-19}$ &$\sim10^{-20}$ \\
$Br(\Upsilon\rightarrow e\mu)$&$\sim10^{-7}$ &$\sim10^{-15}$  &$\sim10^{-17}$ &$\sim10^{-14}$ \\
$Br(\Upsilon(3S)\rightarrow \mu\tau)$&$\sim10^{-7}$ &$\sim10^{-8}$ &$\sim10^{-11}$ &$\sim10^{-9}$ \\
\hline
\end{tabular*}%
\end{table}

{\bf Acknowledgments}

This work is supported by the Major Project of National Natural Science Foundation of China (NNSFC) (No. 12235008), the National Natural Science Foundation of China (NNSFC) (No. 12075074, No. 12075073), the Natural Science Foundation of Hebei province(No. A202201022, No. A2020201002, No. A2023201041), the Natural Science Foundation of Hebei Education Department(No. QN2022173).

\appendix
\section{The couplings}
We discuss some couplings used in this work. We deduce the vertexes of $Z$ boson and two scalar bosons, such as vertexes of $\tilde{e}_{{i}}$-$\tilde{e}^*_{{j}}$-$Z^{{\mu}}$ and $\tilde{\nu}^I_{{i}}$-$\tilde{\nu}^R_{{j}}$-$Z^{{\mu}}$.
\begin{eqnarray}
&&\mathcal{L}_{Z\tilde{e}_{{i}}\tilde{e}^*_{{j}}}=\frac{i}{2}\tilde{e}^*_{{j}} \sum_{a=1}^{3}\Big[(- g_1 \cos{\theta'}_W  \sin\theta_W   + g_2 \cos{\theta'}_W \cos\theta_W + g_{Y X} \sin{\theta'}_W  )Z^{E,*}_{i a} Z_{{j a}}^{E}  \nonumber \\
 &&+\Big(-2 g_1 \cos{\theta'}_W  \sin\theta_W   + (2 g_{Y X}  + g_{X})\sin{\theta'}_W  \Big)Z^{E,*}_{i 3 + a} Z_{{j 3 + a}}^{E}  \Big]\Big(- p^{\tilde{e}^*_{{j}}}_{\mu}  + p^{\tilde{e}_{{i}}}_{\mu}\Big)\tilde{e}_{{i}}Z^{\mu},\nonumber \\
 &&\mathcal{L}_{Z\tilde{\nu}^I_{{i}}\tilde{\nu}^R_{{j}}}=\frac{1}{2}\tilde{\nu}^R_{{j}}\sum_{a=1}^{3} \Big((g_1 \cos{\theta'}_W  \sin\theta_W   + g_2 \cos{\theta'}_W \cos\theta_W  - g_{Y X} \sin{\theta'}_W )Z^{I,*}_{i a} Z^{R,*}_{j a} \nonumber \\
 &&+g_{X} \sin{\theta'}_W Z^{I,*}_{i 3 + a} Z^{R,*}_{j 3 + a}  \Big)\Big(- p^{\nu^R_{{j}}}_{\mu}  + p^{\nu^I_{{i}}}_{\mu}\Big)\tilde{\nu}^I_{{i}}Z^{\mu}.
\end{eqnarray}

The vertexes of $Z$ boson and two fermions are discussed follows, including the vertexes of ${\chi}^+_{{i}}$-${\chi}^-_{{j}}$-$Z^{{\mu}}$, ${\chi}^0_{{i}}$-${\chi}^0_{{j}}$-$Z^{{\mu}}$, $\bar{d}_{{i \alpha}}$-$d_{{j \beta}}$-$Z_{{\mu}}$, $\bar{l}_{{i}}$-$l_{{j}}$-$Z^{{\mu}}$ and $\bar{u}_{{i \alpha}}$-$u_{{j \beta}}$-$Z^{{\mu}}$.
\begin{eqnarray}
&&\mathcal{L}_{Z{\chi}^+_{{i}}{\chi}^-_{{j}}}=\frac{i}{2} {\chi}^-_{{j}}\Big\{\Big[2 g_2  \cos\theta_W  \cos{\theta'}_W  U_{{i 1}}U^*_{j 1} +\Big(- g_1 \cos{\theta'}_W  \sin\theta_W   + g_2 \cos\theta_W  \cos{\theta'}_W  \nonumber \\
 && + (g_{Y X} + g_{X})\sin{\theta'}_W  \Big)U_{{i 2}} U^*_{j 2} \Big]\gamma_{\mu}P_L
 + \Big[2 g_2 \cos\theta_W  \cos{\theta'}_W V^*_{i 1}  V_{{j 1}}+ \Big(- g_1 \cos{\theta'}_W  \sin\theta_W  \nonumber \\
 &&  + g_2 \cos\theta_W  \cos{\theta'}_W   + (g_{Y X} + g_{X})\sin{\theta'}_W  \Big)V^*_{i 2} V_{{j 2}} \Big]\gamma_{\mu}P_R\Big\}{\chi}^+_{{i}}Z^{\mu},\nonumber \\
 &&\mathcal{L}_{Z{\chi}^0_{{i}}{\chi}^0_{{j}}}=-\frac{i}{2}{\chi}^0_{{j}}\Big\{ \Big[ \Big(g_1\cos{\theta'}_W  \sin\theta_W   + g_2 \cos{\theta'}_W \cos\theta_W  - (g_{Y X} + g_{X})\sin{\theta'}_W  \Big)\Big(N_{{i 3}}N^*_{j 3}\nonumber \\
 &&-N_{{i 4}} N^*_{j 4} \Big)\hspace{-0.1cm} -\hspace{-0.1cm}  2 g_{X} \sin{\theta'}_W \Big( N_{{i 6}} N^*_{j 6}\hspace{-0.1cm} -\hspace{-0.1cm} N_{{i 7}} N^*_{j 7} \Big)\Big]\gamma_{\mu}P_L \hspace{-0.1cm} -\hspace{-0.1cm} \Big[ \Big(g_1 \cos{\theta'}_W  \sin\theta_W \hspace{-0.1cm} +\hspace{-0.1cm}  g_2 \cos{\theta'}_W \cos\theta_W \nonumber \\
 && - (g_{Y X} \hspace{-0.1cm} + \hspace{-0.1cm} g_{X})\sin{\theta'}_W  \Big)\Big( N^*_{i 3} N_{{j 3}}\hspace{-0.1cm} - \hspace{-0.1cm} N^*_{i 4} N_{{j 4}}\Big) \hspace{-0.1cm} -\hspace{-0.1cm} 2 g_{X} \sin{\theta'}_W  \Big(N^*_{i 6} N_{{j 6}} \hspace{-0.1cm} -\hspace{-0.1cm} N^*_{i 7} N_{{j 7}} \Big)\Big]\gamma_{\mu}P_R\Big)\Big\}{\chi}^0_{{i}}Z^{\mu},\nonumber \\
 &&
 \mathcal{L}_{Z\bar{d}_{{i \alpha}}d_{{j \beta}}}=\frac{i}{6}\bar{d}_{{i \alpha}}\Big\{ \delta_{\alpha \beta} \delta_{i j} \Big(3 g_2 \cos{\theta'}_W \cos\theta_W  + g_1 \cos{\theta'}_W  \sin\theta_W   - g_{Y X} \sin{\theta'}_W  \Big)\gamma_{\mu}P_L\nonumber \\
 && -\delta_{\alpha \beta} \delta_{i j} \Big(2 g_1 \cos{\theta'}_W  \sin\theta_W   - (2 g_{Y X}  + 3 g_{X} )\sin{\theta'}_W  \Big)\gamma_{\mu}P_R\Big\}d_{{j \beta}}Z^{\mu},\nonumber \\
 &&
 \mathcal{L}_{Z\bar{l}_{{i}}l_{{j}}}=\frac{i}{2}\bar{l}_{{i}}\Big\{ \delta_{i j} \Big(- g_1 \cos{\theta'}_W  \sin\theta_W   + g_2 \cos{\theta'}_W \cos\theta_W  + g_{Y X} \sin{\theta'}_W  \Big)\gamma_{\mu}P_L\nonumber \\
 && -\delta_{i j} \Big(2 g_1 \cos{\theta'}_W  \sin\theta_W   - (2 g_{Y X}  + g_{X})\sin{\theta'}_W  \Big)\gamma_{\mu}P_R\Big\}l_{{j}}Z^{\mu},
 \nonumber \\
 &&
 \mathcal{L}_{Z\bar{u}_{{i \alpha}}u_{{j \beta}}}=-\frac{i}{6} \bar{u}_{{i \alpha}}\Big\{\delta_{\alpha \beta} \delta_{i j} \Big(3 g_2 \cos{\theta'}_W \cos\theta_W  - g_1 \cos{\theta'}_W  \sin\theta_W   + g_{Y X} \sin{\theta'}_W  \Big)\gamma_{\mu}P_L\nonumber \\
 && -\delta_{\alpha \beta} \delta_{i j} \Big(4 g_1\cos{\theta'}_W  \sin\theta_W  - (3 g_{X}  + 4 g_{Y X} )\sin{\theta'}_W  \Big)\gamma_{\mu}P_R\Big\}u_{{j \beta}}Z^{\mu}.
\end{eqnarray}

We also derive the  vertexes of one scalar boson and two different fermions, including vertexes of $\bar{\chi}^-_{{i}}$-$u_{{j \beta}}$-$\tilde{d}^*_{{k \gamma}}$, $\bar{d}_{{i \alpha}}$-${\chi}^-_{{j}}$-$\tilde{u}_{{k \gamma}}$, $\bar{l}_{{i}}$-${\chi}^-_{{j}}$-$\tilde{\nu}^I_{{k}}$, $\bar{l}_{{i}}$-${\chi}^-_{{j}}$-$\tilde{\nu}^R_{{k}}$, $\bar{\chi}^0_{{i}}$-$d_{{j \beta}}$-$\tilde{d}^*_{{k \gamma}}$, $\bar{\chi}^0_{{i}}$-$l_{{j}}$-$\tilde{e}^*_{{k}}$ and $\bar{\chi}^0_{{i}}$-$u_{{j \beta}}$-$\tilde{u}^*_{{k \gamma}}$.
\begin{eqnarray}
&&\mathcal{L}_{\bar{\chi}^-_{{i}}u_{{j \beta}}\tilde{d}^*_{{k \gamma}}}=i\bar{\chi}^-_{{i}}\tilde{d}^*_{{k \gamma}}\delta_{\beta \gamma}\Big[ \Big(- g_2 U^*_{i 1}  Z_{{k j}}^{D}   + Y_{d,{j}}U^*_{i 2} Z_{{k (3 + j)}}^{D}   \Big)P_L+ Y^*_{u,{j}}  V_{{i 2}} Z_{{k j}}^{D} P_R\Big]u_{{j \beta}},
\nonumber \\
 &&\mathcal{L}_{\bar{d}_{{i \alpha}}{\chi}^-_{{j}}\tilde{u}_{{k \gamma}}}=i\bar{d}_{{i \alpha}}\delta_{\alpha \gamma}\Big[ Y_{d,{i}} Z^{U,*}_{k i} U^*_{j 2}P_L+ \Big(- g_2 Z^{U,*}_{k i}  V_{{j 1}}  + Y^*_{u,{i}} Z^{U,*}_{k (3 + i)}  V_{{j 2}} \Big)P_R\Big]{\chi}^-_{{j}}\tilde{u}_{{k \gamma}},
 \nonumber \\
 &&\mathcal{L}_{\bar{l}_{{i}}{\chi}^-_{{j}}\tilde{\nu}^I_{{k}}}= -\frac{1}{\sqrt{2}} \bar{l}_{{i}}\Big(Y_{e,{i}} Z^{I,*}_{k i} U^*_{j 2}P_L-g_2 Z^{I,*}_{k i} V_{{j 1}}  P_R\Big){\chi}^-_{{j}}\tilde{\nu}^I_{{k}},
 \nonumber \\
 &&\mathcal{L}_{\bar{l}_{{i}}{\chi}^-_{{j}}\tilde{\nu}^R_{{k}}}=\frac{i}{\sqrt{2}}\bar{l}_{{i}}\Big( Y_{e,{i}} Z^{R,*}_{k i} U^*_{j 2}P_L-g_2 Z^{R,*}_{k i}  V_{{j 1}} P_R\Big){\chi}^-_{{j}}\tilde{\nu}^R_{{k}},
 \nonumber \\
 &&\mathcal{L}_{\bar{\chi}^0_{{i}}d_{{j \beta}}\tilde{d}^*_{{k \gamma}}}=-\frac{i}{6}\bar{\chi}^0_{{i}}\tilde{d}^*_{{k \gamma}}\delta_{\beta \gamma}\Big\{ \sqrt{2}\Big( g_1 N^*_{i 1}  -3  g_2 N^*_{i 2} +g_{Y X} N^*_{i 5} \Big)Z_{{k j}}^{D} +6 Y_{d,{j}} N^*_{i 3} Z_{{k (3 + j)}}^{D}P_L\nonumber \\
 &&\hspace{1.7cm} + \Big[6 Y^*_{d,{j}} N_{{i 3}}Z_{{k j}}^{D} + \sqrt{2} \Big(2 g_1  N_{{i 1}}  + (2 g_{Y X}  + 3 g_{X} )N_{{i 5}} \Big)Z_{{k (3 + j)}}^{D} \Big]P_R\Big\}d_{{j \beta}},
 \nonumber \\
 &&\mathcal{L}_{\bar{\chi}^0_{{i}}l_{{j}}\tilde{e}^*_{{k}}}=i\bar{\chi}^0_{{i}}\tilde{e}^*_{{k}}\Big\{\Big[ \frac{1}{\sqrt{2}}\Big((g_1 N^*_{i 1} + g_2 N^*_{i 2} +g_{Y X} N^*_{i 5})  Z_{{k j}}^{E} \Big) - Y_{e,{j}} N^*_{i 3} Z_{{k (3 + j)}}^{E}\Big]P_L\nonumber \\
 &&\hspace{1.35cm}-\Big[ \frac{1}{\sqrt{2}}\Big(2 g_1  N_{{i 1}}  + (2 g_{Y X}  + g_{X})N_{{i 5}} \Big)Z_{{k (3 + j)}}^{E} + Y^*_{e,{j}} N_{{i 3}} Z_{{k j}}^{E} \Big]P_R\Big\}l_{{j}},
 \nonumber \\
 &&\mathcal{L}_{\bar{\chi}^0_{{i}}u_{{j \beta}}\tilde{u}^*_{{k \gamma}}}=-\frac{i}{6}\bar{\chi}^0_{{i}}\tilde{u}^*_{{k \gamma}}\delta_{\beta \gamma} \Big\{\Big(\sqrt{2}( g_1 N^*_{i 1} +3 g_2 N^*_{i 2} + g_{Y X} N^*_{i 5} )Z_{{k j}}^{U}+6 Y_{u,{j}} N^*_{i 4} Z_{{k (3 + j)}}^{U}   \Big)P_L\nonumber \\
 &&\hspace{1.7cm} - \Big[-6 Y^*_{u,{j}} N_{{i 4}} Z_{{k j}}^{U} + \sqrt{2}\Big((3 g_{X}  + 4 g_{Y X} )N_{{i 5}} +  4 g_1 N_{{i 1}} \Big)Z_{{k (3 + j)}}^{U} \Big]P_R\Big\}u_{{j \beta}}.
\end{eqnarray}
\section{The hadron matrix elements}
The corresponding hadron matrix elements we used in our work are encoded as:
\begin{eqnarray}
&&\langle0|\bar{v}_V(p_2)\gamma^{\mu}u_V(p_1)|V(p)\rangle=
\frac{f_Vm_V}{N_c}\varepsilon^{\mu*}(p),
\nonumber\\&&\langle0|\bar{v}_V(p_2)\sigma^{\mu\nu}P_Lu_V(p_1)|V(p)\rangle=
-\frac{f_V\epsilon^{\mu\nu p\varepsilon^*(p)}+if_V(p^{\mu}\varepsilon^{\nu*}(p)-p^{\nu}\varepsilon^{\mu*}(p))}{2N_c},
\nonumber\\&&\langle0|\bar{v}_V(p_2)\sigma^{\mu\nu}P_Ru_V(p_1)|V(p)\rangle=
\frac{f_V\epsilon^{\mu\nu p\varepsilon^*(p)}-if_V(p^{\mu}\varepsilon^{\nu*}(p)-p^{\nu}\varepsilon^{\mu*}(p))}{2N_c},
\nonumber\\&&\langle0|\bar{v}_V(p_2)u_V(p_1)|V(p)\rangle\hspace{-0.1cm}=\hspace{-0.1cm}\langle0|\bar{v}_V(p_2)\gamma_5u_V(p_1)|V(p)\rangle\hspace{-0.1cm}=\hspace{-0.1cm}
\langle0|\bar{v}_V(p_2)\gamma^{\mu}\gamma_5u_V(p_1)|V(p)\rangle\hspace{-0.1cm}=\hspace{-0.1cm}0.
\end{eqnarray}
\section{The one-loop functions}
In this section, we give out the corresponding one-loop integral functions, which are read as:
\begin{eqnarray}
&&I_1(x_1,x_2)=\frac{1}{16{\pi}^2}[-(\bigtriangleup+1+\ln{x_{\mu}})+\frac{x_2\ln{x_2}-x_1\ln{x_1}}{(x_2-x_1)}],
\nonumber\\&&I_2(x_1,x_2)=\frac{1}{32{\pi}^2}[\frac{3+2\ln{x_2}}{(x_2-x_1)}-\frac{2x_2+4x_2\ln{x_2}}{(x_2-x_1)^2}
+\frac{2x_2^2\ln{x_2}-2x_1^2\ln{x_1}}{(x_2-x_1)^3}],
\nonumber\\&&I_3(x_1,x_2)=\frac{1}{16{\pi}^2}[\frac{1+\ln{x_2}}{(x_2-x_1)}+\frac{x_1\ln{x_1}-x_2\ln{x_2}}{(x_2-x_1)^2}],
\nonumber\\&&I_4(x_1,x_2)=\frac{1}{96{\pi}^2}[\frac{11+6\ln{x_2}}{(x_2-x_1)}-\frac{15x_2+18x_2\ln{x_2}}{(x_2-x_1)^2}
+\frac{6x_2^2+18x_2^2\ln{x_2}}{(x_2-x_1)^3}\nonumber\\&&\hspace{2.2cm}+\frac{6x_1^3\ln{x_1}-6x_2^3\ln{x_2}}{(x_2-x_1)^4}],
\nonumber\\&&I_5(x_1,x_2)=\frac{1}{16{\pi}^2}[(\bigtriangleup+1+\ln{x_{\mu}})+\frac{x_2+2x_2\ln{x_2}}{(x_1-x_2)}
+\frac{x_2^2\ln{x_2}-x_1^2\ln{x_1}}{(x_2-x_1)^2}],
\nonumber\\&&G_1(x_1,x_2,x_3)=\frac{1}{16{\pi}^2}[\frac{x_1\ln{x_1}}{(x_1-x_2)(x_1-x_3)}+\frac{x_2\ln{x_2}}{(x_2-x_1)(x_2-x_3)}
+\frac{x_3\ln{x_3}}{(x_3-x_1)(x_3-x_2)}],
\nonumber\\&&G_2(x_1,x_2,x_3)=\frac{1}{16{\pi}^2}[-(\bigtriangleup+1+\ln{x_{\mu}})+\frac{x_1^2\ln{x_1}}{(x_1-x_2)(x_1-x_3)}
\nonumber\\&&\hspace{3.0cm}+\frac{x_2^2\ln{x_2}}{(x_2-x_1)(x_2-x_3)}+\frac{x_3^2\ln{x_3}}{(x_3-x_1)(x_3-x_2)}],
\nonumber\\&&J_1(x_1,x_2,x_3,x_4)=\frac{1}{16{\pi}^2}[\frac{x_1^2\ln{x_1}}{(x_1-x_2)(x_1-x_3)(x_1-x_4)}
+\frac{x_2^2\ln{x_2}}{(x_2-x_1)(x_2-x_3)(x_2-x_4)}
\nonumber\\&&+\frac{x_3^2\ln{x_3}}{(x_3-x_1)(x_3-x_2)(x_3-x_4)}+\frac{x_4^2\ln{x_4}}{(x_4-x_1)(x_4-x_2)(x_4-x_3)}],
\nonumber\\&&J_2(x_1,x_2,x_3,x_4)=\frac{1}{16{\pi}^2}[\frac{x_1\ln{x_1}}{(x_1-x_2)(x_1-x_3)(x_1-x_4)}
+\frac{x_2\ln{x_2}}{(x_2-x_1)(x_2-x_3)(x_2-x_4)}
\nonumber\\&&+\frac{x_3\ln{x_3}}{(x_3-x_1)(x_3-x_2)(x_3-x_4)}+\frac{x_4\ln{x_4}}{(x_4-x_1)(x_4-x_2)(x_4-x_3)}],
\end{eqnarray}
with $\Delta=\frac{1}{\epsilon}-r_{\epsilon}+\ln{4\pi}$.

 \end{document}